# Nanowire design by deep learning for energy efficient photonic technologies

Muhammad Usman[1,2,3,*]

[1]*Data61, CSIRO, Clayton 3168 Victoria Australia*

[2]*School of Physics, The University of Melbourne, Parkville 3010 Victoria Australia*

[3]*School of Physics, Monash University, Clayton, 3168 Victoria Australia*

[*]*musman@unimelb.edu.au*

**Abstract:** This work describes our vision and proposal for the design of next generation photonic devices based on custom-designed semiconductor nanowires. The integration of multi-million-atom electronic structure and optical simulations with the supervised machine learning models will pave the way for transformative nanowire-based technologies, offering opportunities for the next generation energy-efficient greener photonics.

## 1. INTRODUCTION

The research on photonic devices is at a crossroads. The science and technology of light generation, detection and conversion is already making a difference in everyday life through many applications such as optical communications, lighting systems, and solar energy harvesting devices. In the Fourth Industrial Revolution (Industry 4.0), disruptive new applications are expected to be advanced through photonics, such as augmented reality and virtual reality (AR/VR), machine vision, medical imaging, lidar, and quantum computing. It is widely anticipated that these emerging photonic technologies will provide profound socio-economic benefits; however, the future of photonic devices hinges on photonics research, which must deliver innovative breakthroughs in nanomaterials to target unique optoelectronic properties while addressing daunting challenges concerning cost effectiveness, ease of manufacturing, and energy efficiency. These are some of the pressing open questions in photonics today that would require scientific breakthroughs in to enable the future of photonics technologies.

In this article, we will discuss our vision and approach towards designing novel nanowire (NW) materials, custom designed with optimal light generation, detection and conversion properties desirable for the advancement of the future photonic technologies. NW materials are highly versatile and workhorse of the photonic industry underpinning many photonic devices such as lasers, photodetectors, and solar cells [1, 2]. The mainstream approach towards NW device design is based on utilising conventional III-V materials (such as GaAs and InP) or silicon, in which the optoelectronic properties are engineered by controlling NW structural parameters (*e.g.,* size and shape). These methods can only offer incremental improvements in the photonic device performance because the NW properties are fundamentally limited by the underpinning material system. However, we propose to design NWs by using a radically different approach in which hitherto scarcely explored III-V materials will be investigated by introducing bismuth (Bi) and nitrogen (N) in the NW design, offering intrinsically unique optoelectronic properties, combined with innovative solutions to overcome the cost and energy efficiency challenges [3, 4, 5]. Further optimisations can be performed by integration with the emerging 2D material substrates [6, 7] and by the surface functionalisation [8, 9].

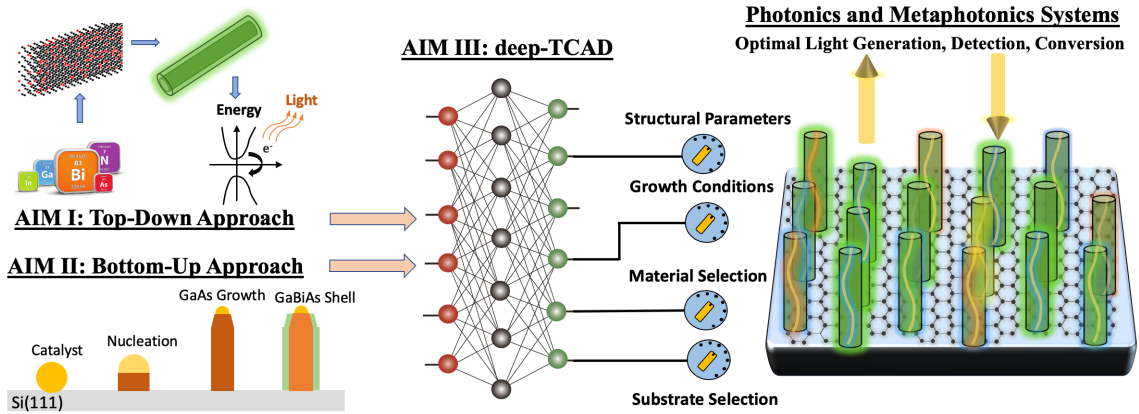

*Figure 1. Schematic diagram illustrating our vision for high-throughput discovery of novel semiconductor nanowire materials which hold the promise for future transformative photonic and metaphotonic technologies.*

We propose that the discovery of novel NWs will be underpinned by the development and application of a new powerful theoretical framework, which will combine advanced machine learning (ML) tools with atomic resolution optoelectronic and growth simulations as illustrated in the *Figure 1*. This will be an exciting, timely and technically demanding opportunity, which will instigate new research directions in the fields of photonic materials and devices. In the proposed approach, NW design will be built upon three foundations, each of which will require innovative developments:

I. **Top-Down Approach:** In this method, we aim to establish a direct quantitative correlation between a nanowire structure and its output optoelectronic and charge transport properties. In this context, multi-million-atom simulations will enable a detailed understanding of the role of nanowire geometry parameters, material profiles, and surface effects.

II. **Bottom-Up Approach:** In this method, we aim to simulate nanowire growth kinetics by performing atomic-resolution molecular dynamic calculations. The crucial new knowledge acquired on the understanding of nanowire growth conditions and their impact on the nanowire morphologies will guide future device fabrication processes.

III. *deep-TCAD* – **Deep Learning Framework:** In this method, we will develop a powerful machine learning framework, *deep-TCAD*, through supervised training of a deep neural network based on the knowledge and data acquired from the above two methods. The trained *deep-TCAD* tool will perform fast, reliable and high-throughput screening of the vast nanowire design space identifying optimal parameters, commensurate with the future photonic technologies.

## 2. LITERATURE BACKGROUND AND OUR PROPOSAL

Semiconductor nanowires (NWs) are highly tuneable nanostructures [1, 2, 10, 11] which provide access to engineered optoelectronic properties tailored for a wide range of devices, including solar cells, photodetectors, laser diodes, light sensors, optical modulators, optical amplifiers, solid-state light sources, and medical imaging [12-15]. Recently, high density of aligned NWs has been fabricated to form new artificial active meta-surfaces suitable for applications in metaphotonic regimes [16]. For the NW growth and design, a large number of material systems have been investigated in the past, including III-Vs [7,8], Si/Ge [9], and ZnO [17]. Among these, the GaAs and the associated alloys are most widely used in the NW design due to the high performing optoelectronic characteristics, the possibility of dislocation free epitaxial growth, and monolithic integration with a large number of substrates including technologically relevant silicon [18] and graphene [6, 7]. However, these materials heavily suffer from the internal non-radiative recombination loss mechanisms [4, 5], which not only severely degrade the device performance, but also heat up the photonic components, requiring additional energy in the form of thermal cooling which in turn significantly increases the operational cost. Therefore, it is highly desirable to pursue new III-V materials which, not only, can offer fundamentally superior electronic and optical properties but can also address the challenges associated with the cost and energy inefficiency of the mainstream III-V materials.

The GaBiAs and its variant materials are an emerging paradigm for photonics with fundamentally unique properties:

*(i)* By varying Bi composition in GaAs material, the band-gap energy can be dramatically modified [4, 5], allowing wavelength tuning over a wide spectral range (1-10 $\mu$m) and making these materials suitable for light emission and sensing devices working in near, mid and far infra-red bands.

*(ii)* The GaBiAs alloys can be designed to increase the spin-orbit splitting energy above the band-gap energy in the telecom spectral range (1.55 $\mu$m) – a property which is not available from any other III-V material or associate alloy system [4, 5]. This unique character of the GaBiAs alloys allows the suppression of the internal loss mechanisms such as inter-valence band absorption (IVBA) and the non-radiative Auger recombination processes (hot holes) [4, 5]. The suppression of these loss mechanisms can drastically reduce the heating up of the photonic devices working in the telecom spectral range, thus eliminating the need for external cooling and significantly reducing the energy consumption and cost [26] of photonic devices with applications in communications and data center sectors.

*(iii)* The incorporation of Bi in photonic devices can reduce their cost due to the lowest cost of Bi among all group III and V elements [3].

These are some of the key inspirations for this work to propose GaBiAs based NW designs which will offer the potential to transform the next generation photonic and metaphotonic technologies.

The incorporation of GaBiAs in NWs is still at a rudimentary stage and the work in this direction has just started to gain traction [19-25]. There are many open questions pertaining the growth, structural characterisation and optoelectronic properties of GaBiAs NWs. In this proposal, the **top-down approach** will develop a highly accurate correlation of NW geometries and compositions with the optoelectronic properties by performing multi-million-atom simulations. In the past, multi-million-atom simulations have been able to quantitatively predict optical and electronic properties of a range of low-dimensional systems such as quantum dots [60-67], quantum wells [30, 31, 59], and impurities in semiconductors [68].

The experimental synthesis of NWs is a very challenging task and the lack of understanding of NW growth kinetics is widely considered a major roadblock in the advancement of NW based photonic technologies. To circumvent this barrier, a quantitative understanding of the NW growth dynamics will be established in the **bottom-up approach**, leading to optimised recipes for the experimental fabrication of NWs incorporating GaBiAs and the related alloys.

In addition to Bi in GaAs material, we will also study another emerging class of nanomaterials by adding Bi in GaNAs materials. These quaternary GaBiNAs materials offer great tuneability of the lattice constant, internal strain, and the optoelectronic properties [27]. In particular, the focus will be to carefully design Bi and N profiles to pursue lattice-matched growth with Ge and/or GaAs substrates, providing high crystalline quality with large volume growth to sandwich GaBiNAs materials as 1 eV absorption layer in tandem InGaP/GaAs/Ge solar cells, offering high solar conversion efficiency and reduced cost [28, 29]. The planned development and application of a **machine learning framework** will create unique opportunities for autonomous, reliable and fast discovery of optimal NW materials and structures, custom designed for the requirements of future transformative photonic and solar energy conversion devices.

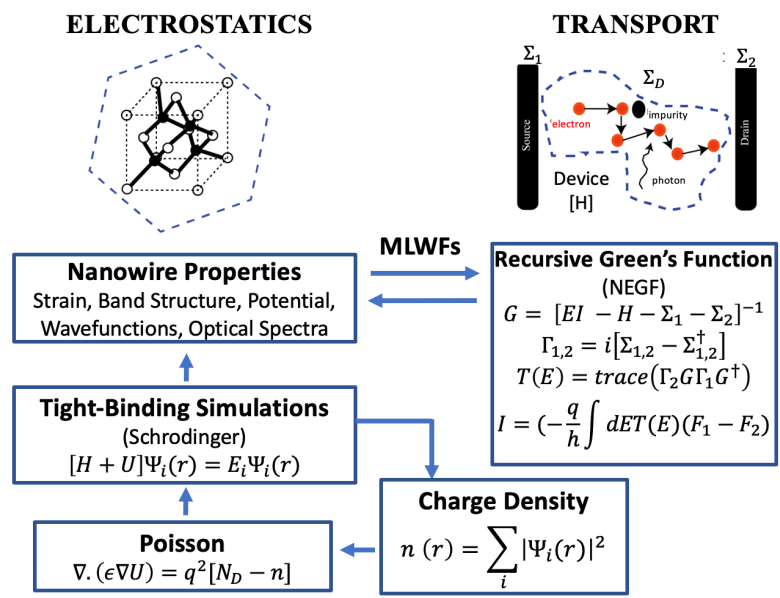

*Figure 2. Schematic diagram of computational framework for nanowire design by top-down approach.*

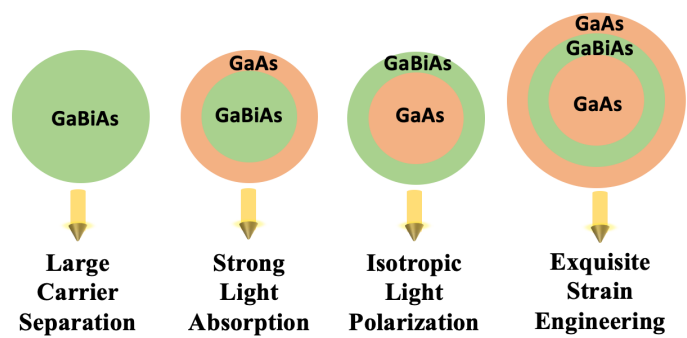

*Figure 3. Nanowire engineering for target functionalities [24, 25].*

## 3. NANOWIRE DESIGN AND METHODOLOGY

Our vision is to develop and apply high-end methodologies to discover new knowledge in emerging areas of research and enable cutting-edge future technologies based on novel NW devices. The proposal has been built upon three major building blocks, each of which are capable of independently generating intriguing new results and instigating innovative research directions: (I) **The top-down approach** based on multi-million-atom optoelectronic simulations will address fundamental open questions pertaining the role of atomic-scale morphology and surface functionalisation of nanowires in defining their optoelectronic and transport character. The answer to these questions will help to identify new nanowire structures with optimal light generation, detection and conversion properties. Notably, this research approach is based on atomistic tight-binding (TB) framework which can include up to 50 million atoms in the simulation domain – a capability which he has established over the last ten years [68]. These state-of-the-art atomistic simulations will provide an unprecedented understanding of the nanowire properties, which is not accessible from the conventional computational methods such as continuum models missing important atomic physics (alloy randomness, surface roughness, etc.), or DFT theory which is restricted to only a few hundred atoms due to the associated high computational costs.

(II) **The bottom-up simulations** will seek to understand highly complex nanowire growth processes involving interaction between many different atoms of various thermodynamic states. The lack of understanding of these processes is considered a major roadblock facing the experimental incorporation of novel GaBiAs and GaBiNAs materials in nanowire devices. The project will also go beyond the mainstream silicon substrate and investigate emerging 2D substrates such as graphene and topological insulators as the nanowire growth platforms, creating new opportunities for the manipulation of their optoelectronic properties.

(III) Another highly innovative part of the proposal is the development of **a powerful machine learning framework**, which will be *first* such development in the world, lowering the computational barrier and automating the nanowire design process.

Our work will perform the design and analysis of advanced nanowire materials by carrying out comprehensive and systematic sets of simulations in both top-down and bottom-up methods. The confluence of these two approaches will empower a machine learning framework, enabling photonic devices with supreme light generation, detection, and conversion efficiencies.

In the following sections, we provide further details about the scientific objectives, methodologies, and the corresponding work-plan in various components of the proposal:

**Top-down Approach.** In this approach, we aim to carry out state-of-the-art atomistic tight-binding simulations to establish a fundamental understanding of the role of nanowire (NW) structural parameters such as size, shape, composition, crystal symmetry, and surface morphology in engineering the optoelectronic and transport properties of photonic devices. *Figure 2* provides a schematic diagram of the computational framework that will be developed and applied to carry out the tasks. The goals in this method will be achieved based on the following four tasks.

**Task 1:** In 2010, we developed a ten-band $sp^3s^*$ tight-binding model to investigate the optoelectronic character of IIIV-Bi-N materials [4, 5, 27]. This model has been very successful to understand the bulk alloy properties in the past ten years [4, 5, 30, 31]. For confined nanostructures like NWs, the capabilities will be further extended by formulating a new set of tight-binding parameters based on extensive 20-band $sp^3d^5s^*$ model with spin. The model will be incorporated in the ELECTROSTATICS component of the theoretical framework which will allow to investigate internal strain, Fermi level, band-structure, wavefunctions, and electron densities as a function of NW geometry parameters such as shape, size, composition profiles and surface morphologies. The work will in particular focus on core-shell and multi-core-shell GaBiAs/GaAs NWs which are the subject of recent experimental developments [19-25]. Our foundational work has shown that these NWs can be engineered to target a variety of optical properties (*Figure 3*). The simulations will be performed over millions of atoms to provide theoretical understanding based on realistic NW sizes ranging over 100 nm. Another parameter of interest is crystal symmetry of NWs. Experiments have shown that GaBiAs NWs can be fabricated in Zincblende (ZB) [21] and Wurtzite (WZ) [20] structures; however, a detailed consequence of these crystal phases for photonic devices is still unknown and will be investigated through a comparative analysis of both ZB and WZ NWs. The work will be performed in close collaboration with experimental groups to confirm the discovered optimised nanowire geometries, enabling photonic device development.

**Task 2:** Nanowires properties can be drastically altered by add-on features including quantum confinement (such as quantum well nanowires) [32, 33] and surface functionalisation [9], which will be the focus of this task. Nanowires with embedded quantum well and quantum dot structures will be studied, which are being actively pursued by research teams for advanced photonic devices such as lasers and LEDs [32, 33]. Atomistic simulations will be carried out to understand the role of nanowire surface functionalisation by the addition of organic and metal-organic functional groups [9] which can expand their applications towards new directions such as biosensing, flexible electronics, and catalysis. To fully exploit the unique capability of GaBiAs materials to suppress energy hungry loss processes, simulations of non-radiative Auger loss mechanisms will also be performed and applied to laser devices [34].

**Task 3:** One of the key issues that hamper solar-to-charge conversion efficiencies in the conventional bulk and thin-film solar cells is very small carrier relaxation times [35, 36]. To address this challenge, this task will couple atomistic simulations with the non-equilibrium Green's function (NEGF) approach to study scattering and transport mechanisms in nanowires, allowing realisation of efficient photo conversion devices such as photodetectors and solar cells. The NEGF approach [37] is a powerful tool that treats quantum transport in nanoscale devices going beyond non-interacting ballistic electronics to include inelastic scattering and strong correlation effects at an atomistic level. *Figure 2* schematically illustrates the connection between atomistic ELECTROSTATIC simulations and NEGF transport in the proposed computational framework. The two models are bridged together via maximally localised Wannier functions (MLWFs) method [38, 39] that allows transforming intrinsically delocalised Bloch functions or plane-waves into localised functions that are essential for the NEGF formalism. Such scheme has been previously implemented in wannier90 [40] and WanT [41] codes but offers scope limited to the ground-state properties and for only up to a few thousands of atoms, not sufficient for nanowire designing. For light conversion devices based on nanowires which are one of the primary targets of this proposal, extension to excited-state dynamics and large-scale simulation domains is required. The ELECTROSTATIC simulations will provide crucial input on strain, band-structures, and carrier confinements to the NEGF TRANSPORT simulations. The theoretical framework proposed in *Figure 2* will offer unique top-down capabilities, on one side providing benefits of multi-million-atom electronic structure calculations including atomistic symmetry effects, random alloy fluctuations, full configuration interaction based many-body theory, *etc*. and on the other hand performing carrier transport including inelastic scattering mechanisms.

**Task 4**: This task will focus on a highly innovative approach towards designing high efficiency tandem solar cell devices by the engineering of GaBiNAs alloys. An example of such devices is multi-junction solar cells (MJ-SC) based on InGaP/GaAs/Ge device architecture [28, 29] which offer the potential for high solar power conversion efficiencies. A key challenge for such triple junction devices is that 1 eV light slips through these cells unabsorbed. For years, scientists have been looking for an *elusive* fourth layer which can fit with InGaP/GaAs/Ge; however, matching atomic structure with GaAs and/or Ge had been a big challenge. GaBiNAs could be that *magical* material, which is cheap, robust, sensitive to 1 eV light, and can be engineered for lattice-matched incorporation as InGaP/GaAs/GaBiNAs/Ge device. This can lead to cost-effective and highly efficient nanowire MJ-SC devices. This task will perform rigorous band structure and charge transport simulations of InGaP/GaAs/GaBiNAs/Ge axial multi-core-shell and superlattice NWs by employing the state-of-the-art theoretical framework developed in tasks 1 and 3, proposing new solar device designs.

**Bottom-up Approach.** Semiconductor nanowires (NW) can be monolithically grown on silicon substrates by the vapor-liquid-solid (VLS) process in molecular beam epitaxy [19-21]. The variations in growth conditions (*temperature, pressure, atomic in-flux*) could lead to drastically different NW structural properties [20, 21]. The past studies have used theoretical simulations [43] and *in-situ* experimental observations [31] to understanding NW growth kinetics; however, the scope is limited to the conventional III-V materials such as GaAs and InP. Despite initial progress on the growth and characterisation of GaBiAs/GaAs NWs [19-25], the impact of Bi incorporation on the GaBiAs and GaBiNAs alloy growth is far from being well understood. Two recent studies have reported drastically different formation of GaBiAs NWs by merely changing the growth temperature, resulting in Zincblende [21] and Wurtzite [20] crystallisations. These NWs offer very different band structure and optical character. The key question to be address here is: whether we can quantitatively understand the role of growth parameters on nanowire synthesis processes to develop reliable recipes, which can guide future experimental developments of NW based photonic devices with target functionalities.

**Task 5:** In 2019, our theoretical research has predicted that about 15% Bi incorporation in GaBiAs/GaAs core-shell NWs is required to target 1.5 µm wavelength for the design of energy efficient telecom photonic devices [24, 25]. So far, only up to 10% Bi has been incorporated in GaBiAs nanowires [48]. One major roadblock is the lack of understanding of NW growth kinetics when Bi and N atoms are incorporated in GaAs. This task will perform systematic and comprehensive set of simulations to understand NW growth dynamics and to formulate a growth recipe for the fabrication of GaBiNAs NWs. The work here will focus on silicon substrates as the NW growth platforms in line with the experimental developments [21]. The target Bi and N compositions will be available from the optoelectronic simulations performed in the method described above. Important questions which will be answered in this task include:

(a) How does growth speed and vapor pressure (or Bi/N atomic flux) depends on substrate material and orientations?

(b) What is the role of catalyst in NW nucleation?

(c) What causes NW growth anomalies such as Bi segregation/clustering and structural kinks?

The answers to these important questions are critical to understand NW growth mechanisms which could guide future experimental development of both axial (core-shell) and superlattice (quantum well) nanowire heterostructures. The work in this task will be performed by carrying out atomistic molecular dynamics (MD) simulations using commercially available LAMMP software package and by developing new Monte Carlo (MC) methods, which have been previously successful for NWs made up of the conventional III-V materials [43]. The application of these advanced atomistic methods will have two-fold impact: firstly, it will go beyond the widely used continuum theories for the analysis of VLS growth which are only good at explaining certain features of NW growth such as the modification of the chemical driving force of growth through the Gibbs–Thomson effect and the dependence of energetically favourable orientations on NW diameter caused by surface energies [47]. Secondly, it will directly provide atomic level details which can be fed back to previous tasks for the investigation of optoelectronic properties based on the multi-million-atom simulations. Our work has recently applied atomistic MD simulations to study the epitaxial growth of SiC on Diamond substrate [49], which can be extended for the work in this task.

**Task 6:** The fabrication of III-V nanowires with the emerging 2D-material substrates such as graphene and topological insulators (TIs) [6, 7, 50] is expected to offer superior electrical, optical, mechanical and transport properties, enabling higher light conversion and detection efficiencies. Contrary to traditional VLS epitaxy widely employed for III-V/Si nanowires, quasi van der Waals epitaxy (QvdWE) technique is emerging as a viable method for the growth of hybrid III-V nanowires/2D-materials devices [50]. Despite recent progress, the growth mechanisms in QvdWE are still not well understood. In this task, Monte Carlo and MD simulations will be performed to develop an understanding of the QvdWE in the context of integration of GaBi(N)As nanowires with graphene and 2D topological insulators.

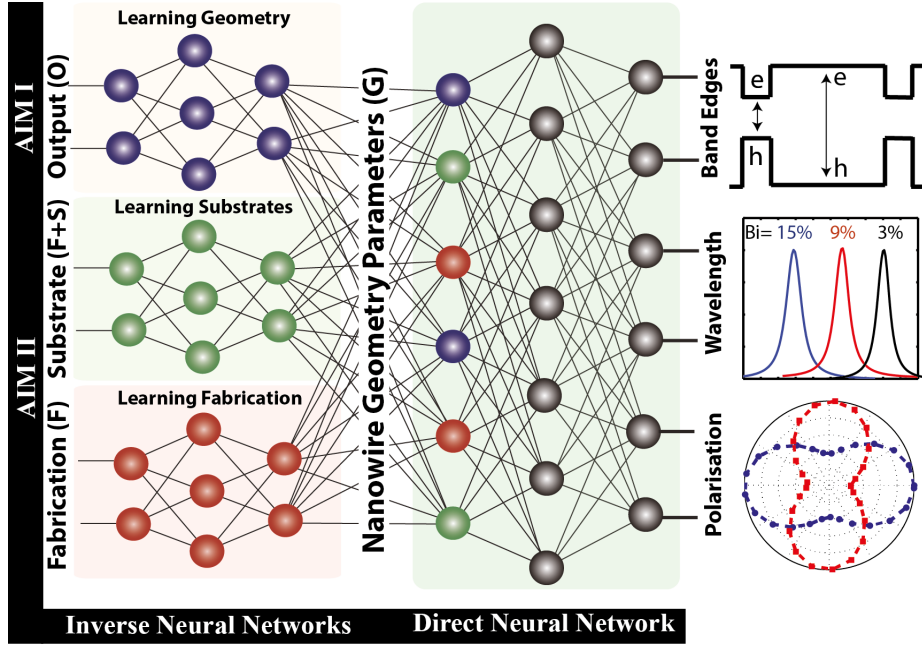

*Figure 4. Schematic of proposed deep-TCAD framework.*

**deep-TCAD – A Machine Learning Framework for NWs:** Materials discovery has long been an expensive pursuit due to its reliance on multiple trial-and-error experiments, expensive reagents and energy-draining manufacturing routes. Even with advancements in density functional theory (DFT) and tight binding (TB) theoretical approaches, the development cycles are slow and inefficient due to the associated heavy computational costs. Contrarily, machine learning (ML) approaches could enable autonomous, robust, time-efficient and accurate discovery of new materials, all at a fractional computational cost, and are therefore receiving a lot of attention lately [51]. Up until recently, ML tools have scarcely been applied to the design of NWs [54, 55] – our proposal will be the *first* to integrate top-down and bottom-up atomistic approaches into a deep learning framework. It is anticipated that the developed ML framework will expedite the nanowire design process by rapidly identifying optimal nanostructures.

*Generation of Data Sets:* A crucial element underpinning the development of the proposed ML capability will be the availability of large datasets and knowledge acquired in tasks 1-6 above.

The variables of interest for the NW design are divided into four sets, with some examples given below (more variables can be added readily):

Geometry: **G** = *{Diameter, Length, III-V composition, Wurtzite, Zincblende, axial structure, superlattice structure, …}*

Output: **O** = *{Strain, Wavelength, Energy band-edges, Polarisation-resolved optical spectra, …}*

Fabrication: **F** = *{Temperature, III/V atomic influx, …}*

Substrate: **S** = *{Silicon, Graphene, 2D TIs, …}*

Three exhaustive datasets (D1-D3) will be generated by changing variables one-by-one from the sets G, O, F, and S above. **D1** will be based on varying G and computing O (tasks 1-4), **D2** will be based on varying F and computing G (task 5), and **D3** will be based on varying F+S and computing G (task 6). The systematic exploration of this vast NW design space will allow the accumulation of several thousand data instances which will supply crucial and sufficient inputs for an artificial intelligence framework to acquire multivariate learning during the *supervised training* of a deep neural network. Smaller but similar datasets, unseen during the training, will evaluate the learning accuracy.

***Deep Learning Architecture:*** In this work, an innovative cascaded deep neural network (CDNN) with fully connected neurons will be trained based on the generated data sets described above. *Figure 4* schematically illustrates the CDNN architecture to be developed in this proposal. Contrary to the traditional approaches employing only a single neural network (NN), the CDNN will be concurrently trained at two levels acquiring multivariate learning of the nonlinear correlations between NW fabrication conditions, geometry parameters and optoelectronic properties, thus gaining the most comprehensive training needed to guide the full assembly-line of NWs. In the first level, three independent NNs will perform the inverse design task, where the training will be based on varying parameters from the sets O, S, and F and mapping them onto the set G. The *blue*, *green* and *red* NNs will take O, S, and F as inputs, respectively, and predict G as output. In the second level, a direct NN will be trained by combining the output neurons of the three inverse NNs into the input neurons. This direct NN will predict back the desired outputs (O) by learning from the geometry parameters (G). The comparison of the final outputs (from direct NN) with the inverse NN inputs will confirm the accuracy of training and learning of the CDNN. Ultimately, the well-trained CDNN framework will be able to take desired output properties (O) and the experimental growth conditions (F+S) and map them onto the actual output parameters, mimicking the growth and measurement of NWs as would be in an actual fabrication set-up. The CDNN will be implemented and trained by using opensource Keras with TensorFlow as the underpinning platform [56, 57]. The project will perform NN optimisations by tuning its parameters such as number of hidden layers, neurons per layer, and number of epochs to achieve optimal learning. The NN will use Adam algorithm with optimised learning rates [58].

***Benchmarking and Discovery:*** The trained CDNN will predict the best possible combinations of G, F and S sets for a desired set O. These predicted optimal NWs will then be rigorously studied by performing full-scale TB simulations to further confirm the accuracy of the CDNN. Additional tuning of the CDNN may be performed by adjusting NN weights to achieve the best match between its predictions with the full TB simulations. We aim to obtain rich amount of experimental data on fabricated NWs from the experimental teams, which will be directly fed to the CDNN to test its accuracy under realistic fabrication environments. At the completion of rigorous training, testing and benchmarking of the CDNN, it will be able to guide future experiments to enable transformative photonic devices. The CDNN will be made available to open-source platform via an easy-to-use web-based interface, which will facilitate nanowire based photonic device design.

## 4. CONCLUSIONS

Nanowires are versatile nanostructures which can be custom designed to enable unique optoelectronic properties. Our proposal describes a comprehensive approach based on our vision for designing novel nanowires, which is based on the state-of-the-art optoelectronic simulations coupled with a supervised machine learning framework. We propose to explore new III-V materials by incorporating Bi and N in GaAs and InP which will enable to target unique properties for the design and implementation of the next generation photonic and electronic devices.